\documentclass[pra,superscriptaddress,showpacs,amstext,amsfont,reprint]{revtex4-1}
\usepackage{amsfonts,amssymb,amsmath,mathrsfs,graphicx,bm,color,appendix}
\usepackage{float}
\usepackage{txfonts}

\begin{document}

\title{Excitation of exciton-polariton vortices in pillar microcavities by a Gaussian beam}
\author{A. S. Abdalla}
\author{Bingsuo Zou}
\affiliation{Beijing Key Lab of Nanophotonics $\&$ Ultrafine Optoelectronic Systems and School of Physics, Beijing Institute of Technology, Beijing 100081, China}%
\author{Yuan Ren}
\author{Tong Liu}
\affiliation{Aerospace Engineering University, School of Astronautics Engineering, Beijing 101416, China}%
\author{Yongyou Zhang}
\email[Author to whom correspondence should be addressed. Electronic mail: ]{yyzhang@bit.edu.cn}
\affiliation{Beijing Key Lab of Nanophotonics $\&$ Ultrafine Optoelectronic Systems and School of Physics, Beijing Institute of Technology, Beijing 100081, China}%

\date{\today}

\begin{abstract}
With coupled Gross-Piteavskii equations we study excitation of exciton-polariton vortices and antivortices in a pillar microcavity by a Gaussian pump beam. The structure of vortices and antivortices shows a strong dependence on the microcavity radius, pump geometry, and nonlinear exciton-exciton interaction. Due to the nonlinear interaction the strong Gaussian beam cannot excite more polariton vortices or antivortices with respect to the weak one. The calculation demonstrates that the weak Gaussian beam can excite vortex-antivortex pairs, vortices with high angular momentum, and superposition states of vortex and antivortex with high opposite angular momentum. The pump geometry for the Gaussian beam to excite these vortex structures are analyzed in detail, which holds a potential application for Sagnac interferometry and generating the optical beams with high angular momentum.

\end{abstract}


\maketitle

\section{Introduction}
Semiconductor microcavities, consisting of two distributed Bragg reflectors, can exhibit spontaneous coherence for exciton polaritons that are bosonic quasiparticles --- a superposition state of excitons in quantum wells and photons in cavities \cite{s1,s2}. The polaritons, due to the photonic part, can be coherently excited by an incident laser and detected by their emitted light \cite{s3,s4,s5,s6,s7}. While the exciton part of the polaritons is responsible for the nonlinear polariton-polariton interaction which have been engineered to produce polariton amplification effects and other spontaneous parametric instabilitites \cite{s8,s9,s10,s11}. Above a pump threshold the polaritons macroscopically occupy the same quantum state, forming a Bose-Einstein condensate \cite{s12,s13}. The polariton condensate attracts major interest because their dispersion, spacial and temporal coherence can be designed by advanced photonic techniques \cite{s15}. As a kind of quantum fluids of light \cite{s17,s18} the polariton condensate has a hydrodynamical-like behavior \cite{s19}, such as superfluidity \cite{s20,s21,s22,s23}, solitons \cite{s24}, quantized vortices \cite{s25,s26}, and structuring of exciton polariton condensates in a pillar microcavities \cite{s27}.  Resonantly pumped polaritons in the optical parametric oscillator regime have been used to show the superfluidity \cite{s28,s29}.

Quantized vortices are topological excitations characterized by the vanishing of the field density at a given point (the vortex core) and the quantized winding of the field phase from 0 to $2\pi l$ ($l$ is a integer) around it \cite{s25,s30}. They have been extensively studied and observed in nonlinear optical systems \cite{s31,s32}, superconductors under magnetic fields \cite{s33,s34}, superfluids \cite{s23,s35}, and cold atoms by setting the system into rotation \cite{s36}. Various vortex states in the polariton condensate continue gaining much attention on disorder effects \cite{s37}, vortex-antivortex pairs \cite{s38,s39,s40}, and vortex ring \cite{s41}. These vortices show a strong dependence on the potential landscape designed by fabrication techniques \cite{s42} or using optical potentials induced by exciton-exciton interactions \cite{s43,s44}. The vortex properties of the non-equilibrium polariton condensates have been diagnosed from experiments \cite{s12,s20,s45,s46,s47} and theories \cite{s17, s58} in last decades, such as lattices of vortices \cite{s49} and superposition of vortex-antivortex states \cite{s50}. To create polariton vortices one can use the Laguerre-Gauss optical beam that carries a well-defined external orbital angular momentum \cite{s51,s52}. The vortex-antivortex superposition states are of potential interest to Sagnac interferometry \cite{s50,s53}, being a gyroscope which has been archived in atomic systems \cite{s54}, and to quantum information \cite{s55,s56}.

\begin{figure}[H]
\center
\includegraphics[width=8.0 cm]{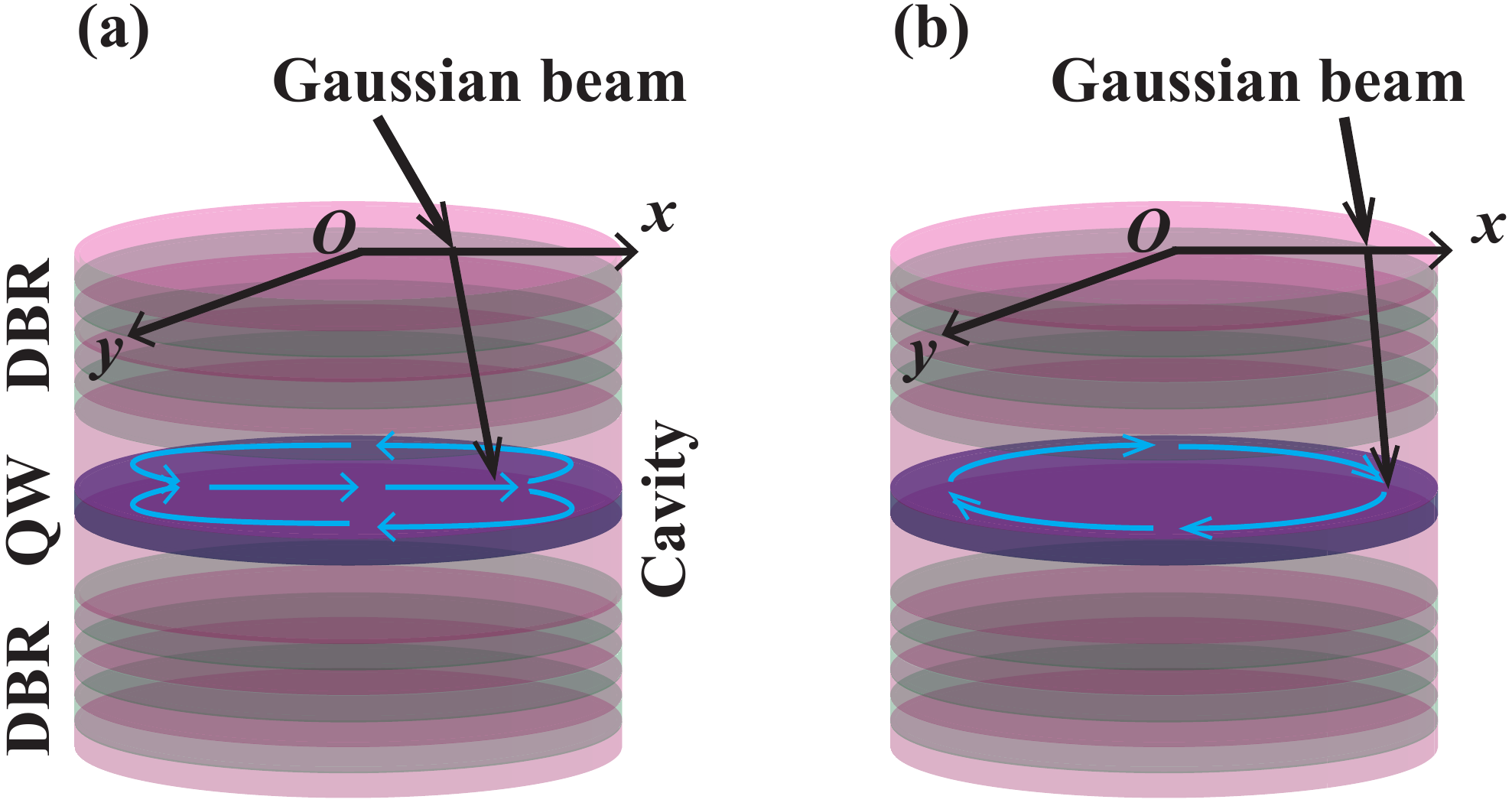}
\caption{Schematic drafts for creating polariton vortices in pillar semiconductor microcavities by one Gaussian beam whose center locates on the $x$ axis and wave vector is (a) along and (b) normal to the $x$ axis. The blue curves show the possible motion path of the polaritons, resulting in polariton vortices or antivortices.}
\end{figure}

The application of the polariton vortices strongly depends on their effective excitation in a semiconductor microcavity, so that we will study how to effectively excite them in the present work. As mentioned above, an efficient method is to use the Laguerre-Gauss beams \cite{s44}. Since more common lasers are Gaussian type, as a rational expectation researchers hope that the vortices could be directly generated by Gaussian beams rather than Laguerre-Gauss beams. Because Gaussian beams do not carry angular momentums, they cannot induce the polariton vortices in an infinite microcavities with translational symmetry. However, we will show that Gaussian beams can excite the polariton vortices in a finite microcavity. In the finite microcavity the geometry of the Gaussian beam and microcavity boundaries play major roles. In addition, the polariton-polariton interaction needs to be considered in the strong pump regime, though it can be neglected in the weak pump regime. Therefore, we will focus on their effects on the excitation of the polariton vortices and antivortices in the present work. Figure 1 shows two possible excitation processes by Gaussian beams. When the polaritons arrive at the microcavity boundary they will change their motion direction and  thus can form the vortices.

The present work is organized as follows. In Sec.~II, we first introduce the coupled dynamic equations for the quantum well excitons and cavity photons from quantum field theory and then give the system parameters adopted in numerical calculation. Numerical results and discussion are shown in Sec.~III which is separated into two subsections according to the pump strength. Finally, a brief conclusion is summarized in Sec. IV.


\section{Hamiltonian and mean-field equations}

The polariton field in a planar microcavity can be described as the coupling of the quantum well exciton field, $\hat \Psi_X(\bm r,t)$, and the cavity photon field, $\hat \Psi_C(\bm r,t)$, and consequently the polariton Halmiltonian is \cite{s17, s58}
\begin{eqnarray}
{\cal H}&=&\int d^2{\bm r}\sum_{i,j}^{\{X,C\}}
\hat{\Psi}_i^\dag({\bm r})\left[\textbf{h}_{ij}^0 +V_{ij}
({\bm r})\delta_{ij}\right]\hat{\Psi}_j({\bm r})\nonumber
\\
&+&{\hbar g_X\over 2}\int d^2{\bm r}\hat{\Psi}_X^\dag({\bm r})
\hat{\Psi} _X^\dag({\bm r})\hat{\Psi}_X({\bm r})\hat{\Psi}
_X({\bm r})\nonumber
\\
&+&\int d^2{\bm r}\hbar F_p({\bm r})
\hat{\Psi}_C^\dag({\bm r})+{\rm H.c}
\end{eqnarray}
where ${\bm r} = (x, y)$ is the in-plane spatial coordinate and the indices $i,j\in\{X,C\}$ denoting the exciton and photon fields, respectively. The field operators for the quantum well excitons and cavity photons satisfy the Bose commutation rules, $[\hat{\Psi}_{i}({\bm r}),\hat{\Psi}^{\dag}_{j}({\bm r'})]=\delta^2({\bm r}-{\bm r'})\delta_{ij}$. The single-particle Hamiltonian, $\textbf{h}^0$, reads
\begin{equation}
\textbf{h}^0=\hbar\left(
\begin{array}{cc}
\omega_X(-i\nabla)&\Omega_R\\
\Omega_R&\omega_C(-i\nabla)
\end{array}\right)
\end{equation}
where the Rabi frequency $\Omega_R$ corresponds to the exciton-photon coupling. The photon dispersion, $\omega_{C}({\bm k})=\omega^{0}_{C}\sqrt{1+{\bm k}^2/k_z^2}$, is a function of the in-plane wavevector, ${\bm k}$, and the quantized photon wavevector in the growth direction, $k_z$. For simplicity, we approximate it to be $\omega_C(\bm k)=\omega_C^0+\frac{\hbar^2 {\bm k}^2}{2m_C}$ with the cavity photon effective mass $m_C$. Because the effective mass is far larger for the excitons than for the cavity photons, we take a flat exciton dispersion, namely, $\omega_{X}({\bm k})=\omega^{0}_X$. In this framework, the polaritons simply arise as the eigenmodes of the linear Hamiltonian in Eq.~(2) and the eigenvalues for the two-branch (upper and lower) polaritons are $\omega_{UP/LP}(\bm k)=\frac{1}{2}\left\{\left[\omega_X^0+\omega_C(\bm k)\right]\pm\sqrt{\left[\omega_X^0-\omega_C(\bm k)\right]^2+4\Omega^{2}_{R}}\right\}$. $V_X(\bm r)$ and $V_C(\bm r)$ in Eq.~(1) are the single particle potentials acting on the exciton and photon fields, respectively. They can break the translational symmetry of the microcavity along the two in-plane directions. The exciton potential generally dates from natural interface or alloy disorder in the quantum wells, while the photon potential is mainly determined by the cavity height or transversal size. Therefore, it is much easier to design the photon potential than to design the exciton potential \cite{s3,s18,s42,s57}. At last, $g_X$ and $F_p({\bm r})$ measure the exciton-exciton interaction and the external pump field, respectively. For convenience, $\hbar$ is set to 1 in the following part if there is no ambiguity.

For solving the polariton system in Eq.~(1), we use the mean-field approximation, namely, $\psi_{X/C}({\bm r})=\langle\hat{\Psi}_{X/C}({\bm r})\rangle$. The mean-field theory has proven to be an efficient way to describe the quantum fluid properties of the polariton condensate. The motion equation of $\psi_{X/C}({\bm r})$, also known as the coupled Gross-Pitaevskii equations \cite{s36,s48}, can be obtained as
 \begin{widetext}
\begin{align}
i\hbar\frac{d}{d t}
\left(\begin{array}{c}
\psi_X({\bm r})\\
\psi_C({\bm r})
\end{array}\right)
=
\left[\mathbf{h}^{0}
+\left(\begin{array}{cc}
-\frac{i}{2}\gamma_X+V_{X}({\bm r})+g_X|\psi_{X}({\bm r})|^{2} & 0\\
0 & -\frac{i}{2}\gamma_C+V_{C}({\bm r})
\end{array}\right)\right]
\left(\begin{array}{c}
\psi_X({\bm r})\\
\psi_C({\bm r})
\end{array}\right)
+
\left(\begin{array}{c}
0\\
F_p({\bm r})
\end{array}\right)
\end{align}
\end{widetext}
from the field Heisenberg motion equation of $i\hbar {d\over dt} {\bf \hat\Psi}(\bm r) = \left[{\bf \hat\Psi}(\bm r),~ ~ {\cal H}\right]$ with ${\bf \hat\Psi}(\bm r)=\left[{\hat\Psi}_X(\bm r), {\hat\Psi}_C(\bm r)\right]^T$. The quantities $\gamma_X$ and $\gamma_C$ are the exciton and photon decay rates, respectively. The Gaussian pump beam considered in the present work is defined as
\begin{align}
F_p(\bm r)= f_pe^{-(\bm r-\bm r_p)^2/w^2}\cdot e^{i\bm k_p\cdot \bm r}\cdot e^{-i\omega_{p}t}
\end{align}
where $f_p$, $w$, and $\omega_p$ denote the amplitude, spot size, and frequency of the pump field, respectively. $\bm r_p=(x_p,y_p)$ is the center coordinate of the pump spot. The pump wave vector, $\bm k_p=(k_{px},k_{py})$, can be adjusted by the incident angle of the pump field with respect to the growth direction. The incident strength of the Gaussian beam is proportional to $|f_p|^2$. When $\bm k_p\neq 0$ the excited polaritons have a non-zero flow velocity along the cavity plane and therefore, it is possible for them to form the quantum vortices.

Without loss of generality, we consider a pillar microcavity as shown in Fig.~1. For a pillar microcavity $V_C(\bm r)$ is infinite outside the cavity region due to the total internal reflection on the boundaries \cite{s48}. In calculation $V_C(\bm r)=100$ meV for $r>R$ and 0 for others, which cuts out the required pillar microcavity with radius $R$. Note that we mainly focus on the effects of the pump geometry in the present work and therefore, $V_X(\bm r)$ is set to zero to avoid the disorder influence.

In the following numerical calculation the parameters of a typical GaAs-based microcavity are adopted. The energy of the excitons is taken as the zero point, i.e., $\omega^{0}_X=0$, and other parameters are $m_C=1\times 10^{-5}m_e$ where $m_e$ is the free electron mass, $\gamma_X=\gamma_C=0.01$ meV, $g_X=0.015$ ${\rm meV}\cdot\mu {\rm m}^2$ \cite{s58}, $\Omega_R=2.5$ meV. In addition, two pump cases, namely, weak pump regime with $f_p=0.1$ ${\rm meV}\cdot\mu {\rm m}^{-1}$ and another strong with $f_p=100$ ${\rm meV}\cdot\mu {\rm m}^{-1}$, are considered to show the influence of the nonlinear interaction on the polariton vortices. As is well known the exciton part concerns the nonlinear interaction, while the photon part in the polaritons relates to the pump efficiency. Consequently, the polaritons with $\bm k=\bm k_p$ had better have a suitable ratio between them, for example, half over half. This requires $\omega_C(\bm k_p) = \omega_X^0$, always maintained in the following calculation. Besides, we take the pump detuning to be $\delta_p=\omega_p-\omega_{LP}(\bm k_p)=-0.2$ meV and the Gaussian beam size to be $w=4$ $\mu m$.

\begin{figure}
\center
\includegraphics[width=7 cm]{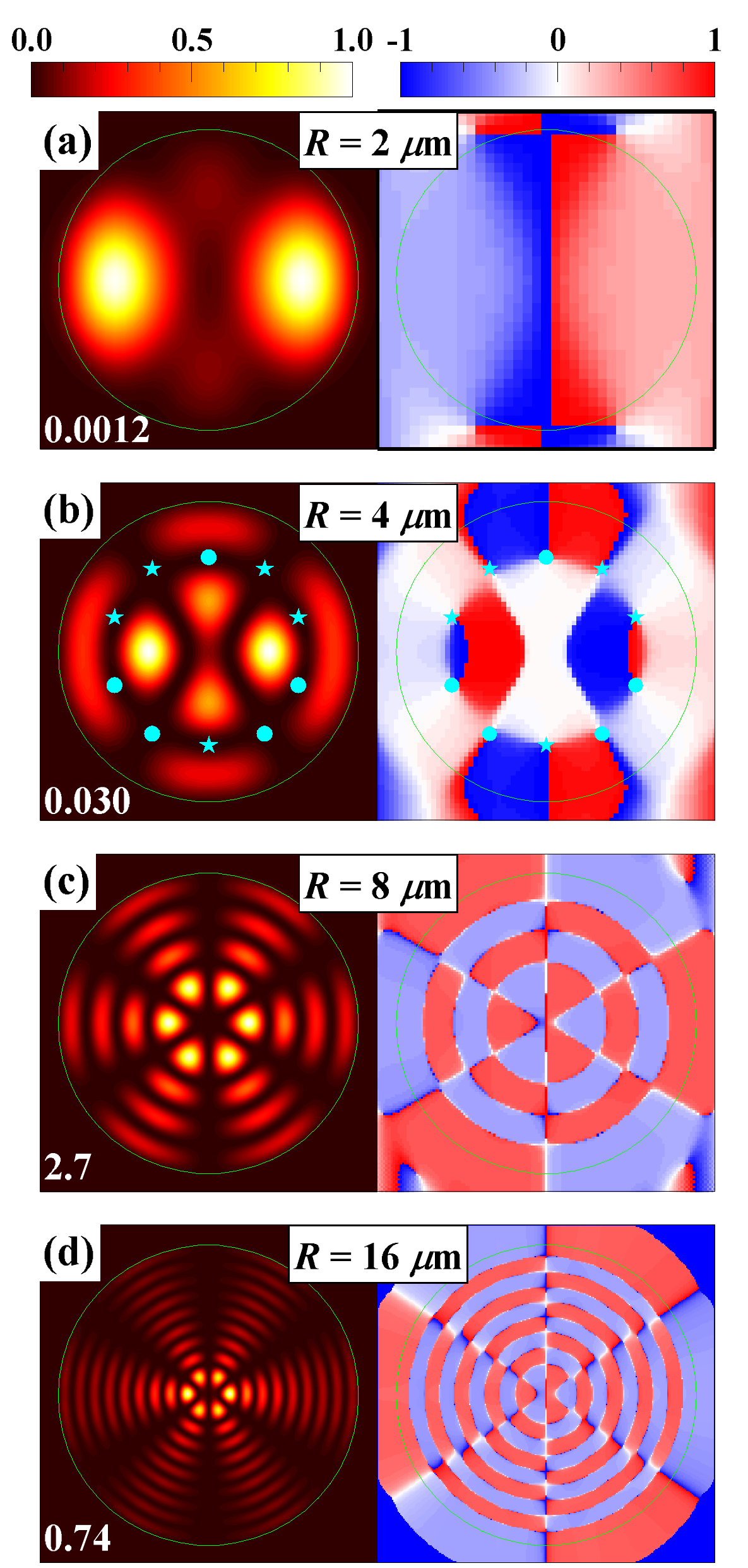}
\caption{Steady density distribution of photon fields, $|\psi_C(\bm r)|^2$, in the left column and corresponding phase (with unit of $\pi$) distribution in the right column. The numbers in the bottom left corner denote the maximum value of the photon density and the green circles represent the pillar boundaries with radii (a) $R = 2\ \mu {\rm m}$, (b) $R = 4\ \mu {\rm m}$, (c) $R = 8\ \mu {\rm m}$, and (d) $R = 16\ \mu {\rm m}$. As an example, the vortices and antivortices are denoted by dots and stars in (b). Other parameters: $f_p=0.1\ {\rm meV}\cdot\mu {\rm m}^{-1}$, $\bm k_p=(2,\ 0) \ \mu {\rm m}^{-1}$, $\bm r_p = (0,\ 0)$, and $w=4\ \mu {\rm m}$.}
\end{figure}

\section{Numerical results and discussion}

The polariton superfluid has been generated by several types of pump fields \cite{s12,s18,s44}. In the present work we use the ``resonant injection" scheme that the pump frequency $\omega_p$ is set to be near the lower-branch polariton energy at the pump wave vector, $\omega_{LP}(\bm k_p)$. The Gaussian beam creates the polariton condensate and determines its properties (such as momentum, energy, density, phase). This controllable scheme allows to study the excitation of vortices and antivortices \cite{s39,s40}. For clear we divide the present section into two subsections according to the pump field strength: (A) weak pump regime and (B) strong pump regime. For the former the exciton density is low and thus the nonlinear exciton-exciton interaction can be neglected, while for the later the exciton density is so high that the nonlinear interaction must be considered.

The main results are obtained by numerically solving Eq.~(3) on a two-dimensional grid $320\times320$ for a square $32\ \mu{\rm m}\times32\ \mu{\rm m}$ microcavity region. The discretization area is $0.1\ \mu{\rm m}\times0.1\ \mu{\rm m}$, smaller than the requirement of the maximum pump wave vector $4.0\ \mu{\rm m}^{-1}$ adopted in calculation. The fourth-order Runge-Kutta algorithm is used to evaluate the photon and exciton fields $\psi_{C/X}(\bm r, t)$.

\begin{figure*}
\begin{minipage}{12.5 cm}
\flushleft
\includegraphics[width=12.0 cm]{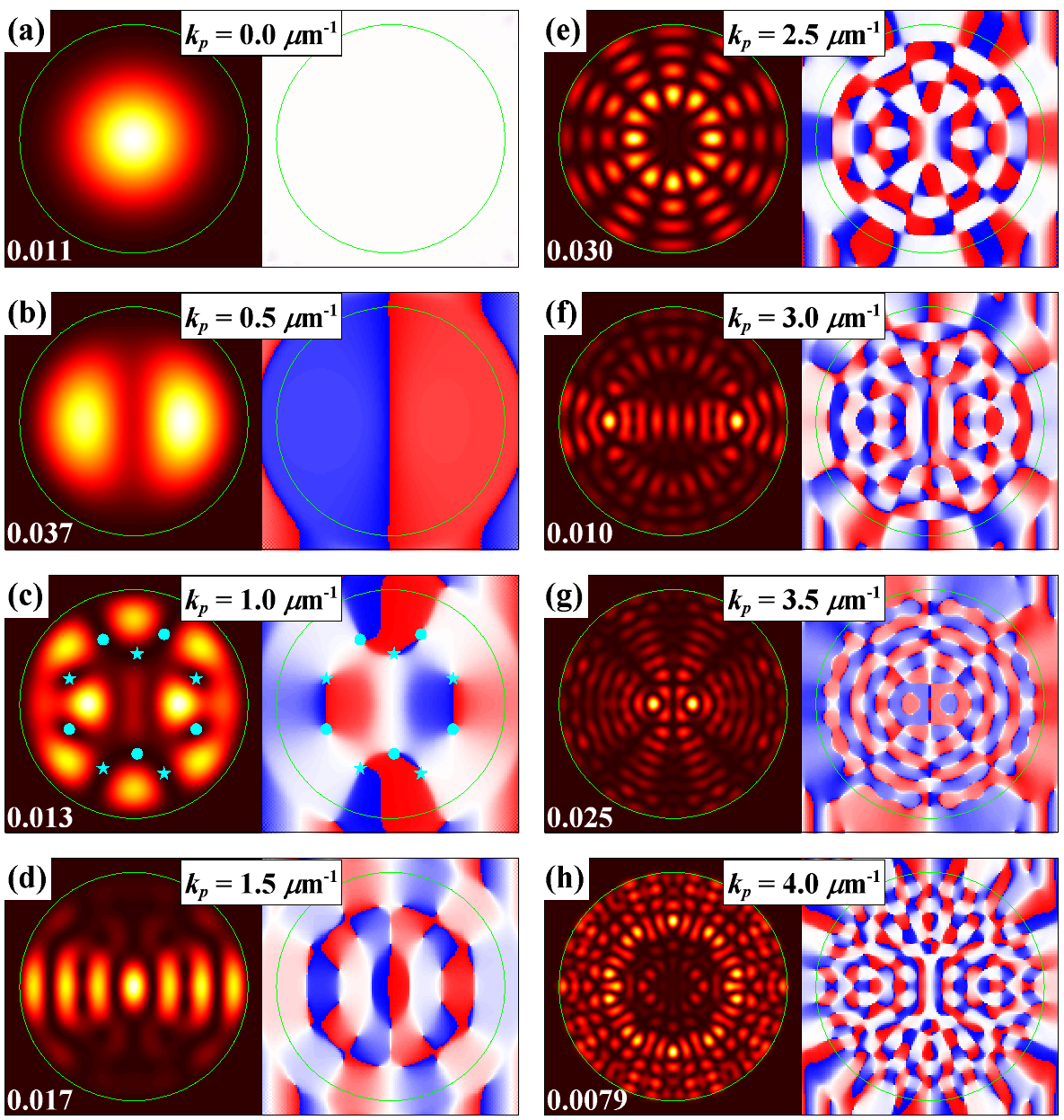}
\end{minipage}
\begin{minipage}{4.8 cm}
\flushright
\caption{Steady density distributions of photon fields, $|\psi_C(x,y,t)|^2$ (left panel in each subfigure) and  corresponding phase distributions (right part). Color scale bars are the same to Fig.~2. The number in the bottom left corner represents the maximum value of the photon density and all green circles denoting the microcavity boundaries have the same radii $R=8\ \mu$m. The wave vector of the pump beam, $\bm k_p$, is along the $x$ axis and its value given in the above of each figure. As an example, the vortices and antivortices are shown by dots and stars in (c). Other parameters: $f_p=0.1\ {\rm meV}\cdot\mu {\rm m}^{-1}$, $\bm r_p = (0,\ 0)$, and $w=4\ \mu {\rm m}$.}
\end{minipage}
\end{figure*}

\subsection{Weak pump regime}

In the weak pump regime the energy due to the exciton-exciton interaction, $g_X|\psi_X(\bm r)|^2$, is far less than the polariton kinetic energy and therefore, its effect can be neglected and the polariton evolution can reach a steady state. Figure 2 shows the steady density distributions of the photon field, $|\psi_C(\bm r)|^2$, in the first column and corresponding field phases in the second column for four microcavities with radii $R=2\ \mu$m, $4\ \mu$m, $8\ \mu$m, and $16\ \mu$m. The exciton field has a similar distribution and so is not shown. The typical velocity of the polariton is ${d\omega_{LP}(\bm k)\over d\bm k}$ and therefore, the characteristic length $\xi\sim {d\omega_{LP}(\bm k)\over d\bm k}\cdot{\hbar\over \gamma_c}$. When $k_p=2\ \mu{\rm m}^{-1}$, $\xi\sim150\ \mu{\rm m}$ and subsequently the polaritons in the four microcavities shown in Fig.~2 can reach the boundary. Due to $R$ less than half of $150\ \mu{\rm m}$ the boundary exerts manifest influence on the polariton condensate in four cases. The width of the Gaussian pump beam with center at the original point is $w=4\ \mu$m, thus with increasing $R$ the microcavity boundary is increasingly away from the pump beam. The pump beam covers all the microcavity when $R\lesssim w$ [see Figs.~2(a-b)], while only a central part when $R>w$ [see Fig.~2(d)]. Accounting for the loss of the polaritons in traveling, the boundary plays a major role on the forming of the polariton states for the small microcavities, that is, the boundary effect decreases with increasing $R$. This can be seen from the variation of the photon density distribution from Figs.~2(a) to 2(d).

As is well known the Gaussian beam has no orbital angular momentum and so cannot excite the vortex by itself. For the present circular and no disorder microcavities to excite the vortices requires two conditions: (i) the microcavity boundary can influence the movement of the polaritons and (ii) the Gaussian pump beam has a nonzero in-plane wave vector. As illustrated in Fig.~1, the polaritons change their moving direction once they arrive at the boundary, accompanied by a complex polariton interference. The polariton interference leads to different spatial structures for the vortices and antivortices, see Figs.~2 and 3. In Fig.~2 the Gaussian beam has a wave vector $\bm k_p=(2,\ 0)\ \mu{\rm m}^{-1}$ and therefore, can induce the vortices and antivortices. For example, the vortices and antivortices denoted by dots and stars in Fig.~2(b). The numbers of the dots and stars are same, due to the mirror symmetry of the pump beam along direction $x$. From Fig.~2(b) the distance between two adjacent vortices can be estimated to be $\sim 1.8\ \mu$m (about half of the polariton wavelength) for $k_p=2\ \mu{\rm m}^{-1}$. Therefore, it is impossible to generate the vortex excitation for small enough microcavities, also called as photonic dots \cite{s58}. Since $R=2\ \mu{\rm m}>1.8\ \mu{\rm m}$ in Fig.~2(a), the superposition state of the vortex and antivortex with $l=\pm1$ is generated. With increasing $R$ more complex polariton vortices can be excited [see Fig.~2(b)], even the vortices with high angular momentum [see Figs.~2(c-d)]. The photon density distributions in Figs.~2(c-d) represent the superposition state of the vortex and antivortex with $l=\pm3$, which is important for application of polaritons to the Sagnac interferometry \cite{s50}. The Sagnac interferometry requires large $l$ whose value is mainly determined by $\bm k_p$ in the present pump geometry.

\begin{figure*}
\center
\includegraphics[width=17.5 cm]{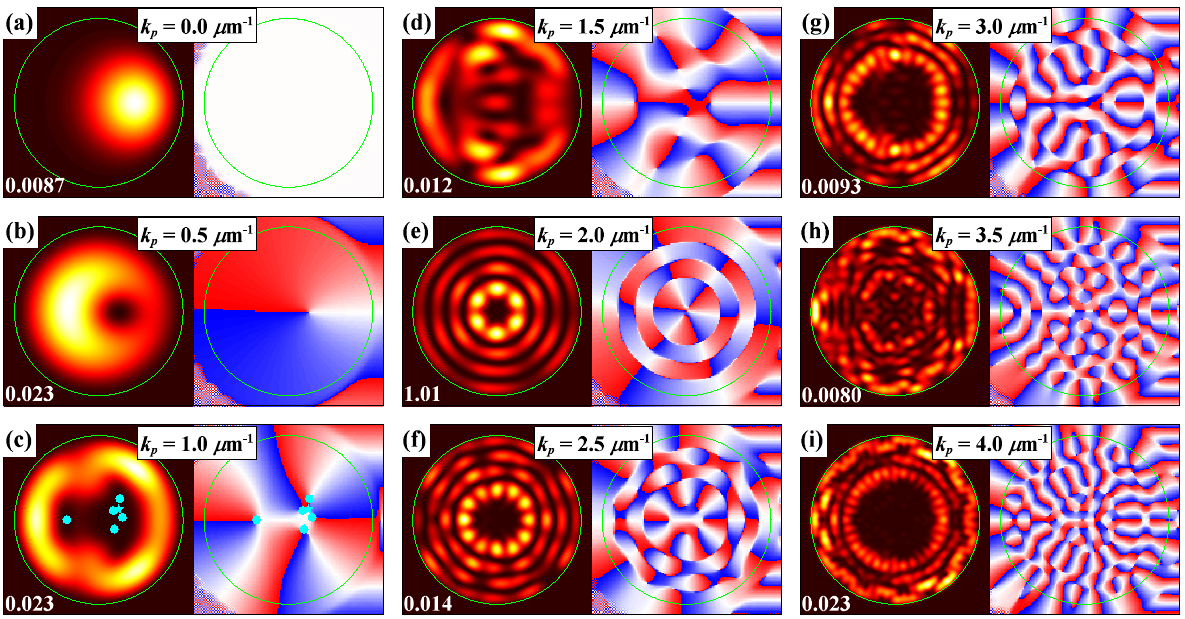}
\caption{Steady density distributions of photon fields, $|\psi_C(x,y,t)|^2$ (left panel in each subfigure) and  corresponding phase distributions (right part). Color scale bars are the same to Fig.~2. The number in the bottom left corner represents the maximum value of the photon density and all green circles denoting the microcavity boundaries have the same radii $R=8\ \mu$m. The wave vector of the pump beam, $\bm k_p$, is along the $y$ axis and its value is given in the above of each figure. As an example, the vortices and antivortices are shown by dots and stars in (c). Other parameters: $f_p=0.1\ {\rm meV}\cdot\mu {\rm m}^{-1}$, $\bm r_p = (4,\ 0)\ \mu {\rm m}$, and $w=4\ \mu {\rm m}$.}
\end{figure*}

The variation of the vortex excitation with $\bm k_p$ is shown in Fig.~3 where the pillar radii are set to $R=8\ \mu{\rm m}$. When $\bm k_p$ is small [Figs.~3(a)] no vortex or antivortex is excited, while with increasing $\bm k_p$ the pattern of the photon density shows more and more complexity and subsequently the vortex and antivortex structures are generated [Figs.~3(b-h)]. In other words, the argument that the distance between two adjacent vortices decreases with increasing $k_p$ is responsible for the complicated density patterns of the high angular momentum states in the large-$k_p$ cases. The high angular momentum states with $|l|>1$ are not energetically favored in a Bose-Einstein condensation, and so they commonly break up into several vortices with $l=\pm1$ \cite{s36}, as shown in Figs.~3(b-d). However, the superposition state of the vortex and antivortex with $l=\pm3$ in Fig.~2(c) is ultra stable, indicating that for a certain pump geometry the polariton condensation in the pillar microcavity can have high angular momentum. Other examples are those superposition states in Fig.~3(e) with $l=\pm6$ and in Fig.~3(h) with $l=\pm16$. Since $\bm k_p$ is along the $x$ axis in Fig.~3, there is a mirror symmetry for the vortices and antivortices along the $x$ axis, referred to Fig.~3(c). This mirror symmetry can be broken up by changing the pump geometry from Fig.~1(a) to Fig.~1(b).

We take the pump geometry of Fig.~1(b) in Fig,~4 where the pump position is set to $\bm r_p=(4,\ 0)\ \mu{\rm m}$ and the wave vector $\bm k_p$ is along the $y$ axis. Since this pump beam has a non-zero angular momentum with respect to the center of the pillar microcavity,  the number of the vortices is different from that of the antivortices. The angular momentum of the pump field  is
\begin{align}
L_{\rm pump} = -i\hbar{\langle F(\bm r)|{\partial_\phi}|F(\bm r)\rangle
\over \langle F(\bm r)|F(\bm r)\rangle} = \bm r_p\times\hbar \bm k_p
\end{align}
where $\phi$ is the azimuth angle of $\bm r$. When $\bm r_p=(0,0)$ there is no net angular momentum for polaritons, i.e., the cases shown in Fig.~3. On the contrary, for a nonzero $\bm r_p$ the net angular momentum of the polariton condensates should be proportional to $\bm k_p$, see Fig.~4 where the total angular momentum is 0, $1\hbar$, $4\hbar$, $6\hbar$, $3\hbar$, $6\hbar$, $11\hbar$, $15\hbar$, and $20\hbar$ from Figs.~4(a) to 4(i), respectively. Because the angular momentum is not conserved, these values are not exactly equal to $L_{\rm pump} $, but maintain the approximate proportional relation with $\bm k_p$.

\begin{figure*}
\begin{minipage}{12.5 cm}
\flushleft
\includegraphics[width=12.0 cm]{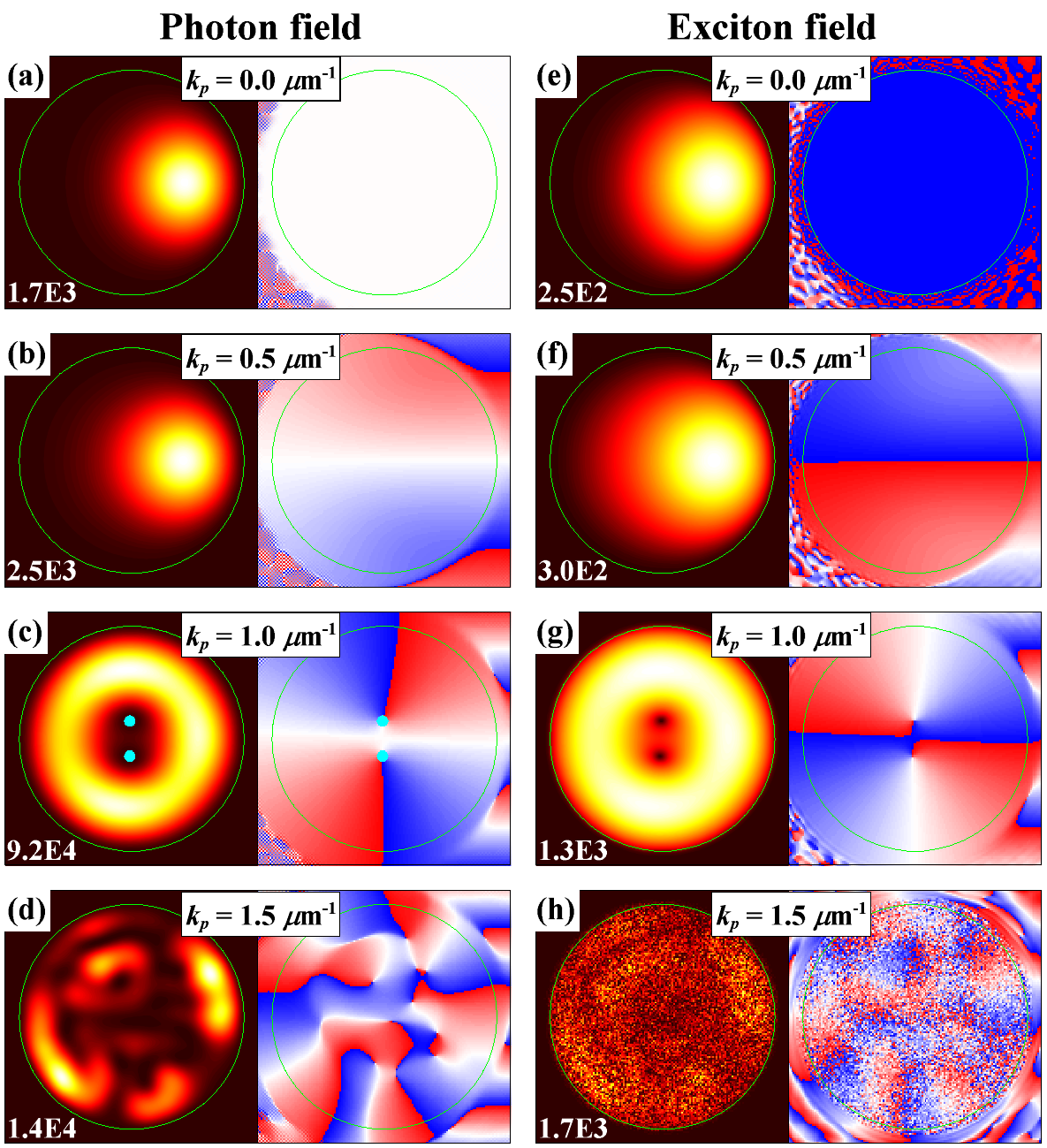}
\end{minipage}
\begin{minipage}{4.8 cm}
\flushright
\caption{Density and phase distributions for (a-d) photon fields and (e-h) exciton fields. The left and right panels in each subfigure show the density and phase distributions, respectively. The distributions in (a-c) and (e-f) are steady, while those in (d) and (h) at the evolution time of $1600\ \hbar\cdot{\rm meV}^{-1}$ are not. Color scale bars are the same to Fig.~2. The number in the bottom left corner represents the maximum value of the photon or excition density and all green circles denoting that the microcavity boundaries have the same radii $R=8\ \mu$m. The wave vector of the pump beam, $\bm k_p$, is along the $y$ axis and its value is given in the above of each subfigure. As an example, the vortices are shown by dots in (c) and (g). Other parameters: $f_p=100\ {\rm meV}\cdot\mu {\rm m}^{-1}$, $\bm r_p = (4,\ 0)\ \mu{\rm m}$, and $w=4\ \mu {\rm m}$.}
\end{minipage}
\end{figure*}

For convenient analysis we expand the photon field, $\psi_C(\bm r)$, as follow
\begin{align}
\psi_C({\bm r})=\sum_{n,l}C_{n,l}\phi_{n,l}(\bm r)
\end{align}
where $\phi_{n,l}(\bm r)$ is a basis function with the angular momentum of $l\hbar$. $n$ and $l$ are the radial and angular quantum numbers, respectively. In the weak pump regime the coefficient $C_{n,l}$ for the steady states can be found with the approximation of neglecting the nonlinear exciton-exciton interaction, i.e.,
 \begin{align}
C_{n,l} &= {\left[\omega_p-\left(\omega_X^0-\frac{i}{2}\gamma_X\right)\right]f_{n,l}e^{-i\omega_pt}\over \left[\omega_p-\left(\omega_C^0+\omega_{n,l}-\frac{i}{2}\gamma_C\right)\right]\left[\omega_p-\left(\omega_X^0-\frac{i}{2}\gamma_X\right)\right]-\Omega^2}.
\end{align}
where $f_{n,l}$ is the pump coefficient for the state $\phi_{n,l}(\bm r)$. The expression of $\phi_{n,l}(\bm r)$ and more information about Eqs.~(6) and (7) are shown in appendix. The total angular momentum for the photon field, $L_{C}$, is given by
 \begin{align}
L_{C}=\hbar{\sum_{n,l}l|C_{n,l}|^2\over\sum_{n,l}|C_{n,l}|^2}.
\end{align}
According to Eq.~(7), the angular momentum of the photon field is mainly determined by the pump coefficient $f_{n,l}$ and corresponding energy $\omega_{n,l}$. For example, the states of $\phi_{0,0}$, $\phi_{0,1}$, $\phi_{0,4}$, and $\phi_{3,3}$ dominate in Figs.~4(a), 4(b), 4(c), and 4(e), respectively. In addition, even the photon fields have the same total angular momentum, their density distribution could be much different from each other, as shown in Figs.~4(d) and 4(f). This is due to that in Fig.~4(d) the states of $\phi_{1,6}$ and $\phi_{2,6}$ dominate while in Fig.~4(f) $\phi_{3,6}$ dominates. When $\bm k_p$ is large the photon density distribution appears more complex, see Figs.~4(g-i) where more than one states of $\phi_{n,l}$ are excited. If one want to excite only one angular momentum state $\phi_{n,l}$ [see Figs.~2(c), 3(e), and 4(e)], equation (7) provides a guidance: (i) increase $f_{n,l}$ by controlling the pump geometry and (ii) achieve a resonant excitation for $\phi_{n,l}$ by tuning $\omega_p$. To summarize, by designing the pump geometry the Gaussian beam can efficiently excite the polariton vortices and antivortices in the pillar microcavity, which holds potential applications for Sagnac interferometry and optical beams with high angular momentum.

\subsection{Strong pump regime}

When the pump field is strong enough the nonlinear exciton-exciton interaction plays an important role in exciting the polariton vortices and antivortices, and can make the steady state of the polaritons unreachable. Figure 5 shows the density and phase distribution for the photon and exciton fields under a strong pump field $f_p=100\ {\rm meV}\cdot\mu {\rm m}^{-1}$. The strong pump field leads to the exciton energy being much higher than $\omega_p$, so that the photon density is much larger than the exciton density, comparing Figs.~5(a-d) with Figs.~5(e-h), respectively. This also can be seen from Eqs.~(A.8) and (A.9): $C_{n,l}\gg X_{n,l}$ if the exciton energy $\omega_X^0$ is much larger than $\omega_p$. Note that in the weak pump regime the densities for the photon and exciton fields are in the same order of magnitude. For small $\bm k_p$ [see Figs.~5(a-c) and 5(e-g)] the polariton system can reach a steady state, while for large $\bm k_p$ the polariton system is unstable [see Figs.~5(d) and 5(h)]. This is owed to that the case with $\bm k_p=1.5\ \mu{\rm m}^{-1}$ displayed by Figs.~5(d) and 5(h) has the largest exciton density, as well as the strongest nonlinear effect. Due to the large value of $g_X|\psi_X(\bm r)|^2$ the temporally excited vortices make the exciton field at different position out of step and subsequently the exciton field appears a random density/phase distribution, referred to Fig.~5(h). This kind of randomness cannot be seen from the photon field, because it is much stronger than the exciton field.

For the steady cases in Figs.~5(a-c) the total angular momentums, respectively, are 0, 0, and $2\hbar$. With respect to Figs.~4(b-c) the figures 5(b-c) hold less total angular momentum, indicating that the strong pump cannot excite more steady vortices. This can be argued as follow: the nonlinear interaction prefers to spread the polariton density equally, while the vortices and antivortices have zero-density points. In addition, since it is not easy to observe the vortices or antivortices in an unstable state, to achieve a steady polariton condensate implies that a not-too-strong pump beam is a better choice.

\section{conclusion}

We have studied the excitation of exciton polariton vortices and antivortices in the pillar microcavities by Gaussian pump beams and found that the structure of vortices and antivortices are strongly dependent on the microcavity radius, pump geometry, and nonlinear exciton-exciton interactions. These parameters for one Gaussian beam to excite the vortices and antivortices are analyzed in detail. We show that it is hard to observe the excited polariton vortices in the strong pump regime because the nonlinear exciton-exciton interaction prefers to spread the polariton density equally and can cause the system to be unstable. On the contrary the polariton system can reach a steady state in the weak pump regime. We show that though the Gaussian pump beams do not carry angular momentums, they can also excite many kinds of the vortex and antivortex structures in the pillar microcavities, such as vortices with high angular momentum, and superposition states of vortex and antivortex with high opposite angular momentum. Our results demonstrate that exciting vortices and antivortices by Gaussian beams are possible for experimental observation, which holds potential applications for Sagnac interferometry, quantum information, and generating the optical beams with high angular momentum.

\section*{Acknowledgements}
This work is supported by NSFC (Grant No. 11304015) and Beijing Higher Education Young Elite Teacher Project (Grant No. YETP1228).

\appendix*
\section{Steady solution for Eq.~(3)}

In the present work the pillar microcavity can be taken as an infinite potential well for polaritons. Therefore, it is convenient to expand the fields $\psi_X(\bm r)$ and $\psi_C(\bm r)$ into
\begin{align}
\left(\begin{array}{c}
\psi_X({\bm r})\\
\psi_C({\bm r})
\end{array}\right)
=
\sum_{n,l}\left(\begin{array}{c}
X_{n,l}\\
C_{n,l}
\end{array}\right)
\phi_{n,l}(\bm r)
\end{align}
where the basis function $\phi_{n,l}(\bm r)$ satisfies
\begin{align}
\left[-{\hbar^2\over 2m_C}\nabla^2+V_C(\bm r)\right]\phi_{n,l}(\bm r)=\hbar\omega_{n,l}\phi_{n,l}(\bm r).
\end{align}
$\phi_{n,l}(\bm r)$ takes the form
\begin{align}
\phi_{n,l}(\bm r)=J_{n,|l|}(r)\times{1\over\sqrt{2\pi}}e^{il\varphi}
\end{align}
where $n$ and $l$ represent the radial and angular quantum numbers, respectively. $J_{n,|l|}(r)=J_{|l|}(k_nr)$ is a normalized Bessel function with boundary condition $J_{|l|}(k_nR)=0$ and corresponding energy $\hbar\omega_{n,l}={\hbar^2k_n^2\over 2m_C}$.
Similar to $\psi_X(\bm r)$ and $\psi_C(\bm r)$, we also expand $F_p(\bm r)$ as
\begin{align}
F_p(\bm r)
=
\sum_{n,l}
f_{n,l}
\phi_{n,l}(\bm r)e^{-i\omega_pt}.
\end{align}
where
\begin{align}
f_{n,l}
=
f_p\int d^2\bm r \phi_{n,l}^*(\bm r)e^{-(\bm r-\bm r_p)^2/w^2} e^{i\bm k_p\cdot \bm r}.
\end{align}
When $\bm r_p$ and $\bm k_p$, respectively, are on and along the $x$ axis, we have $f_{n,l}=f_{n,-l}$. Substituting Eqs.~(A.1) and (A.5) into Eq.~(3), one obtains the dynamical equations for $C_{n,l}$ and $X_{n,l}$ as follows
 \begin{widetext}
 \begin{subequations}
 \begin{eqnarray}
i\hbar{d\over dt}C_{n,l} &=& \left(\omega_C^0+\omega_{n,l}-\frac{i}{2}\gamma_C\right)C_{n,l}+\Omega X_{n,l}+f_{n,l}e^{-i\omega_pt},\\
i\hbar{d\over dt}X_{n,l} &=& \left(\omega_X^0-\frac{i}{2}\gamma_X\right)X_{n,l}+\Omega C_{n,l}+g_X\sum_{n_1n_2n_3,l_1l_2l_3}X_{n_1l_1}X_{n_2l_2}^*X_{n_3l_3}\int d^2\bm r\phi_{n,l}^*({\bm r})\phi_{n_1l_1}({\bm r})\phi_{n_2l_2}^*({\bm r})\phi_{n_3l_3}({\bm r}).
\end{eqnarray}
\end{subequations}
\end{widetext}
For the weak pump, the nonlinear interaction term $g_X$ can be neglected, and subsequently the equation (A.6) for the steady state reads
 \begin{subequations}
 \begin{eqnarray}
\omega_p C_{n,l} &=& \left(\omega_C^0+\omega_{n,l}-\frac{i}{2}\gamma_C\right)C_{n,l}+\Omega X_{n,l}+f_{n,l}e^{-i\omega_pt},\nonumber\\
\omega_p X_{n,l} &=& \left(\omega_X^0-\frac{i}{2}\gamma_X\right)X_{n,l}+\Omega C_{n,l}.\nonumber
\end{eqnarray}
\end{subequations}
From above equations, one can find the $C_{n,l}$ and $X_{n,l}$ for the steady state as follow
 \begin{align}
C_{n,l} &= {\left[\omega_p-\left(\omega_X^0-\frac{i}{2}\gamma_X\right)\right]f_{n,l}e^{-i\omega_pt}\over \left[\omega_p-\left(\omega_C^0+\omega_{n,l}-\frac{i}{2}\gamma_C\right)\right]\left[\omega_p-\left(\omega_X^0-\frac{i}{2}\gamma_X\right)\right]-\Omega^2},\\
X_{n,l} &={\Omega f_{n,l}e^{-i\omega_pt}\over \left[\omega_p-\left(\omega_C^0+\omega_{n,l}-\frac{i}{2}\gamma_C\right)\right]\left[\omega_p-\left(\omega_X^0-\frac{i}{2}\gamma_X\right)\right]-\Omega^2}.
\end{align}
Substituting Eqs.~(A.8) and (A.9) into Eq.~(A.1), one can directly obtain the steady state of the polariton system in the weak pump regime. On the other hand, the nonlinear term should be considered in the strong pump regime and leads to that the dynamical equation cannot reach a steady state in common. As a result it is hard to observe the excited polariton vortices or antivortices for the strong pump. Since the nonlinear interaction in the strong pump regime makes the exciton energy much higher than $\omega_p$, the density of the photon field is much higher than that of the exciton field.

\end{document}